\documentclass{article}

\pdfoutput=1
\usepackage{arxiv}

\usepackage[utf8]{inputenc} 
\usepackage[T1]{fontenc}    
\usepackage{hyperref}       
\usepackage{url}            
\usepackage{booktabs}       
\usepackage{amsfonts}       
\usepackage{nicefrac}       
\usepackage{microtype}      
\usepackage{lipsum}
\usepackage{graphicx}
\usepackage{amsmath}
\usepackage{subcaption}
\usepackage{graphicx}
\graphicspath{ {./images/} }

\title{Dual-Modality Computational Ophthalmic Imaging with Deep Learning and Coaxial Optical Design}

\author{
  Boyuan Peng\footnotemark[1] \\
  Shenzhen International Graduate School\\
  Tsinghua University\\
  \texttt{pengby99@outlook.com}
  \And
  Jiaju Chen\thanks{Equal contribution.} \\
  Shenzhen International Graduate School\\
  Tsinghua University\\
  \texttt{chen-jj22@mails.tsinghua.edu.cn}
  \And
  Yiwei Zhang\footnotemark[1] \\
  Shenzhen International Graduate School\\
  Tsinghua University\\
  \texttt{1342732139@qq.com}
  \And
  Cuiyi Peng \\
  Shenzhen International Graduate School\\
  Tsinghua University\\
  \texttt{pengcy22@mails.tsinghua.edu.cn}
  \And
  Junyang Li \\
  Shenzhen International Graduate School\\
  Tsinghua University\\
  \texttt{lijunyang22@mails.tsinghua.edu.cn}
  \And
  Jiaming Deng \\
  Shenzhen International Graduate School\\
  Tsinghua University\\
  \texttt{}
  \And
  Peiwu Qin\thanks{Corresponding author.} \\
  Shenzhen International Graduate School\\
  Tsinghua University\\
  \texttt{pwqin@sz.tsinghua.edu.cn}
}

\begin{document}
\maketitle
\begin{abstract}
The growing burden of myopia and retinal diseases necessitates more accessible and efficient eye screening solutions. This study presents a compact, dual-function optical device that integrates fundus photography and refractive error detection into a unified platform. The system features a coaxial optical design using dichroic mirrors to separate wavelength-dependent imaging paths, enabling simultaneous alignment of fundus and refraction modules. A Dense-U-Net-based algorithm with customized loss functions is employed for accurate pupil segmentation, facilitating automated alignment and focusing. Experimental evaluations demonstrate the system’s capability to achieve high-precision pupil localization (EDE = 2.8 px, mIoU = 0.931) and reliable refractive estimation with a mean absolute error below 5\%. Despite limitations due to commercial lens components, the proposed framework offers a promising solution for rapid, intelligent, and scalable ophthalmic screening, particularly suitable for community health settings.
\end{abstract}


\section{Introduction}
Refraction testing is essential for diagnosing and correcting refractive errors, which occur when incoming light fails to properly focus on the retina. Using tools such as manual or automated refractors, clinicians can assess refractive power and astigmatism, indirectly evaluating lens deformation and guiding interventions such as prescription correction\cite{1}. Fundus examination is a key non-invasive method for evaluating the posterior segment of the eye, particularly the retina, optic nerve, and vasculature. Techniques such as direct or indirect ophthalmoscopy are widely used to detect conditions like diabetic retinopathy, macular degeneration, and retinal detachment. Advances in fundus photography have greatly improved the ability to capture detailed retinal images, enhancing the accuracy of ocular disease diagnosis and monitoring\cite{2}.

Refraction measurement analyzes optical reflections from the eye to estimate refractive errors, with outcomes influenced by individual variations in ocular geometry and light paths. The principle was first introduced by Cuignot in 1873 using retinal reflection for retinoscopy, laying the foundation for objective refraction assessment\cite{3}. Later developments, such as eccentric photorefraction by Kaakinen, expanded image features by capturing both corneal and retinal reflections, though they introduced angular distortion\cite{4}. With advances in image processing, modern systems achieve automated and accurate refraction analysis from single images, using techniques like edge detection, pupil fitting, and optical modeling\cite{5}. Recent methods based on wavefront aberrometry and OCT further enhance accuracy by capturing high-order aberrations, though their complexity and cost limit routine clinical use\cite{6, 7}.

Fundus imaging is essential for evaluating posterior eye structures and has evolved significantly since Helmholtz introduced the direct ophthalmoscope in 1851\cite{8}. Indirect ophthalmoscopes later expanded the field of view and reduced aberrations through aspheric lens design\cite{9}. Modern fundus cameras, originating in the 1920s, have transitioned from mechanical systems to compact, optoelectronic devices. Internal illumination has become the mainstream design due to its superior light efficiency and image clarity\cite{10, 11}. Recent innovations include non-mydriatic infrared imaging, portable systems, and scanning laser techniques to enhance field coverage and reduce artifacts\cite{12, 13}.

Given the high clinical value of ocular examinations, there is a growing need to extend such services beyond large medical centers, which often face limitations in accessibility. In practice, multiple tests across different rooms and operators reduce both spatial and temporal efficiency. To address this, we propose a compact, integrated device that combines multiple eye examinations within a single platform. Among potential modalities, refraction testing is more precise and less dependent on patient cooperation than visual acuity tests, while fundus photography offers a straightforward assessment of retinal morphology and is more suitable for primary screening than OCT, which is better suited for detailed follow-up in diagnosed cases. By integrating refraction and fundus imaging into one system, and incorporating automation and AI-assisted analysis, the device can enhance usability in community settings, reduce operator dependency, and enable early detection of conditions such as myopia and diabetic retinopathy.

In summary, our key contributions include:
\begin{itemize}
\item We proposed an integrated optical path design that combines refraction and fundus imaging by analyzing and fusing the respective optical systems, enabling unified imaging and system-level optimization.
\item We developed an image processing approach that enhances the quality and accuracy of fused imaging results through improvements in both optical design and software-based correction.
\item We implemented an automated workflow that reduces manual intervention by optimizing the human-machine interface and enabling fully automatic eye positioning, focusing, and image capture through real-time image analysis.
\end{itemize}

\section{Methods}
\label{sec:headings}
As discussed previously, this study adopts visible-light fundus photography and open-view refraction as the core modalities, with a focus on integrating their respective optical systems. Since the two methods utilize different illumination wavelengths and optical characteristics, their distinction can be leveraged to design a combined optical path. Currently, no existing work has successfully integrated the optical paths of fundus photography and refraction measurement. The main challenges lie in optical compatibility and practical usability. Refraction systems are optimized for precise diopter measurements—such as myopia, hyperopia, and astigmatism—requiring narrow field imaging and accurate focus. In contrast, fundus cameras are designed for wide-field, high-resolution imaging of the retina. These differences in focal requirements and imaging geometry pose significant integration difficulties. Moreover, the two modalities demand different patient positioning and eye conditions. Fundus imaging typically requires pupil dilation, while refraction tests are often performed under ambient light conditions. These differences necessitate flexible illumination control and ergonomic design. Additionally, both systems are traditionally bulky, and direct combination could result in excessive size, limiting clinical applicability. To address these challenges, compact, modular optomechanical design principles are adopted in the integration process, as detailed in the following sections.
\paragraph{Optical Design of Fundus Photography and Refraction Systems.}
Since the retina does not emit light, external illumination is essential for fundus photography. However, strong reflections from the cornea can obscure retinal features, reducing image contrast. Therefore, the illumination system must deliver sufficient light to the retina while minimizing anterior segment reflections. Modern CCDs with broad spectral sensitivity enable multi-wavelength illumination. In visible light, pupil constriction occurs, necessitating pharmacological dilation, which causes patient discomfort and delays. To avoid this, a non-mydriatic strategy is adopted: near-infrared light is used for alignment and low-light viewing, while yellow-green light is used during image acquisition.

The optical design of the fundus imaging path includes four key components: imaging, illumination, positioning, and focusing. The current system emphasizes the first two; positioning and focusing are primarily handled via image processing algorithms, discussed in later sections. The imaging system forms a conjugate relationship with the retina and employs a sensor responsive to both ~550 nm (yellow-green) and ~740 nm (NIR) wavelengths. A triple-filter wheel selects the appropriate bandpass filter for each phase. Kohler illumination is used to achieve uniform light distribution and high-contrast imaging, which is particularly beneficial for the low-reflectivity retinal surface. To reduce stray reflections from the corneal surface, an annular aperture is placed in the illumination path, suppressing near-axial glare and improving image contrast and edge sharpness. This annular light pattern also assists in eye positioning. Additionally, a central black dot mask is inserted into the imaging channel to block direct light from entering the sensor and to prevent ghost images caused by internal reflections. An internal focusing mechanism is employed to compensate for refractive errors, enabling the system to adaptively maintain conjugate focus with the retina across different refractive states. The illumination and imaging systems share a coaxial design to ensure alignment with the visual axis.

Similar to the fundus imaging system, the optical path for open-view refraction consists of an illumination and imaging system. The illumination system includes a natural-light fixation target to relax accommodation and an incident NIR beam for measurement. The imaging system captures the retinal reflection of the NIR light after it passes through the eye’s refractive media. Unlike fundus photography, which requires broad, diffuse illumination, the refraction system uses a narrow-divergence NIR beam. To suppress central corneal reflections and facilitate ring-pattern image formation, an annular aperture is again employed. A focusing system is also integrated to account for refractive errors, ensuring the reflected ring image is properly projected onto the camera sensor. Figure 1 shows the whole construction for fundus imaging and open-view refraction optical paths.

\begin{figure}[ht]
    \centering
    \begin{subfigure}[b]{\textwidth}
        \centering
        \includegraphics[width=0.8\textwidth]{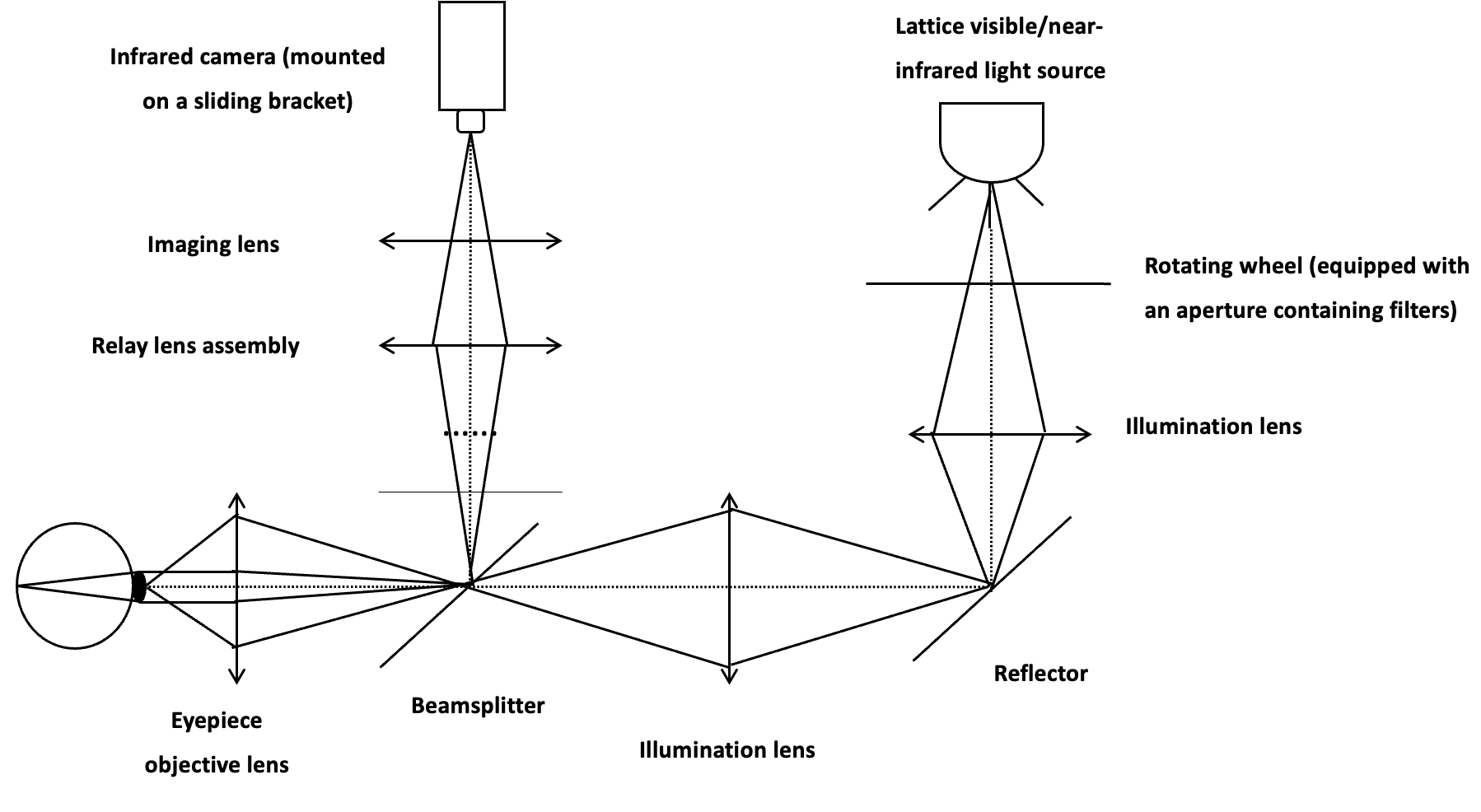}
        \caption{Fundus imaging}
        \label{fig:fig1a}
    \end{subfigure}

    \vspace{1em}

    \begin{subfigure}[b]{\textwidth}
        \centering
        \includegraphics[width=0.8\textwidth]{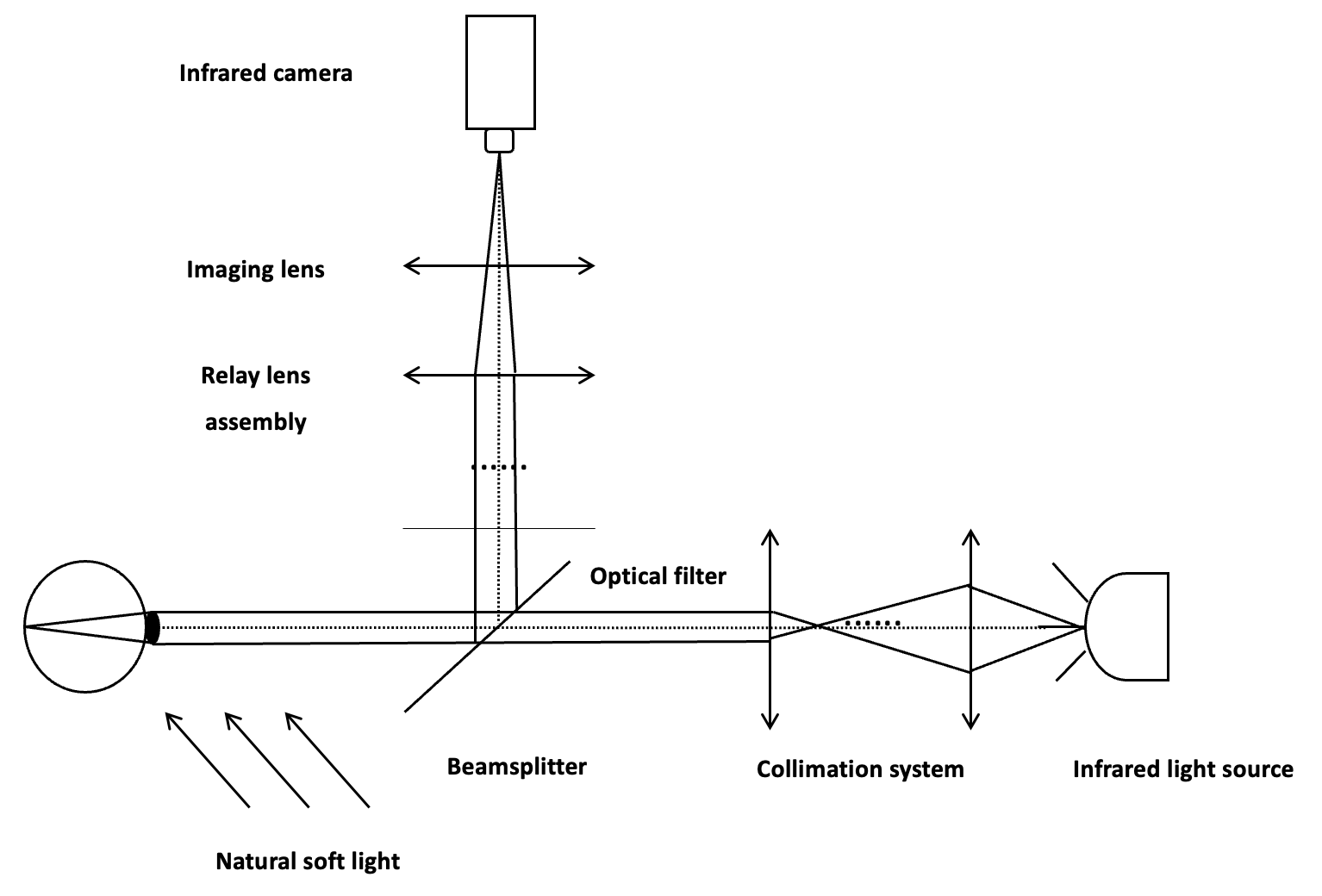}
        \caption{Refraction optical paths}
        \label{fig:fig1b}
    \end{subfigure}

    \caption{Construction of system; (a) Fundus imaging; (b) Refraction optical paths}
    \label{fig:fig1}
\end{figure}
\paragraph{Integrated Optical Design for Combined Fundus Imaging and Refraction.}
As described in the previous sections, the imaging principles of fundus photography and open-view refraction are largely similar, differing mainly in their illumination strategies and the need for relaxed accommodation during refraction. To integrate both systems, a dichroic mirror is used to separate the two illumination paths based on wavelength. A long-pass dichroic mirror with a transmission range of 830–1300nm and a reflection range of 400–760 nm was selected. This configuration effectively separates the 740 nm near-infrared light used for fundus illumination from the 980 nm infrared light used for refraction. Additionally, the ring-shaped aperture used in fundus illumination is combined with a diffusion sheet to produce soft, collimated light, simulating distant illumination and promoting eye relaxation.

To ensure a compact and coaxial optical layout, a folded design inspired by OCT systems is adopted. The fundus imaging and refraction illumination paths are arranged horizontally on the first layer, while the fundus illumination and refraction imaging paths are arranged vertically on a second layer. This design not only reduces the device footprint but also minimizes stray light interference between subsystems.

Focusing is achieved using two stepper motors with a positioning accuracy of 0.01 mm. The subject’s head is stabilized with a chin rest, which is laterally adjustable to align the eye with the optical axis. A broadband industrial camera (MV-CS013-60GN, 1224×1024 resolution) is used for both imaging modules to ensure hardware consistency. However, due to reduced sensitivity at 980 nm (approximately 28\% of peak response), the beam splitter and eyepiece mount used for fundus imaging are rigidly fixed together and detached during refraction measurements to avoid excessive signal loss caused by multiple optical reflections.

\begin{figure}[ht]
    \centering
    \begin{subfigure}[b]{0.45\textwidth}
        \centering
        \includegraphics[width=\textwidth]{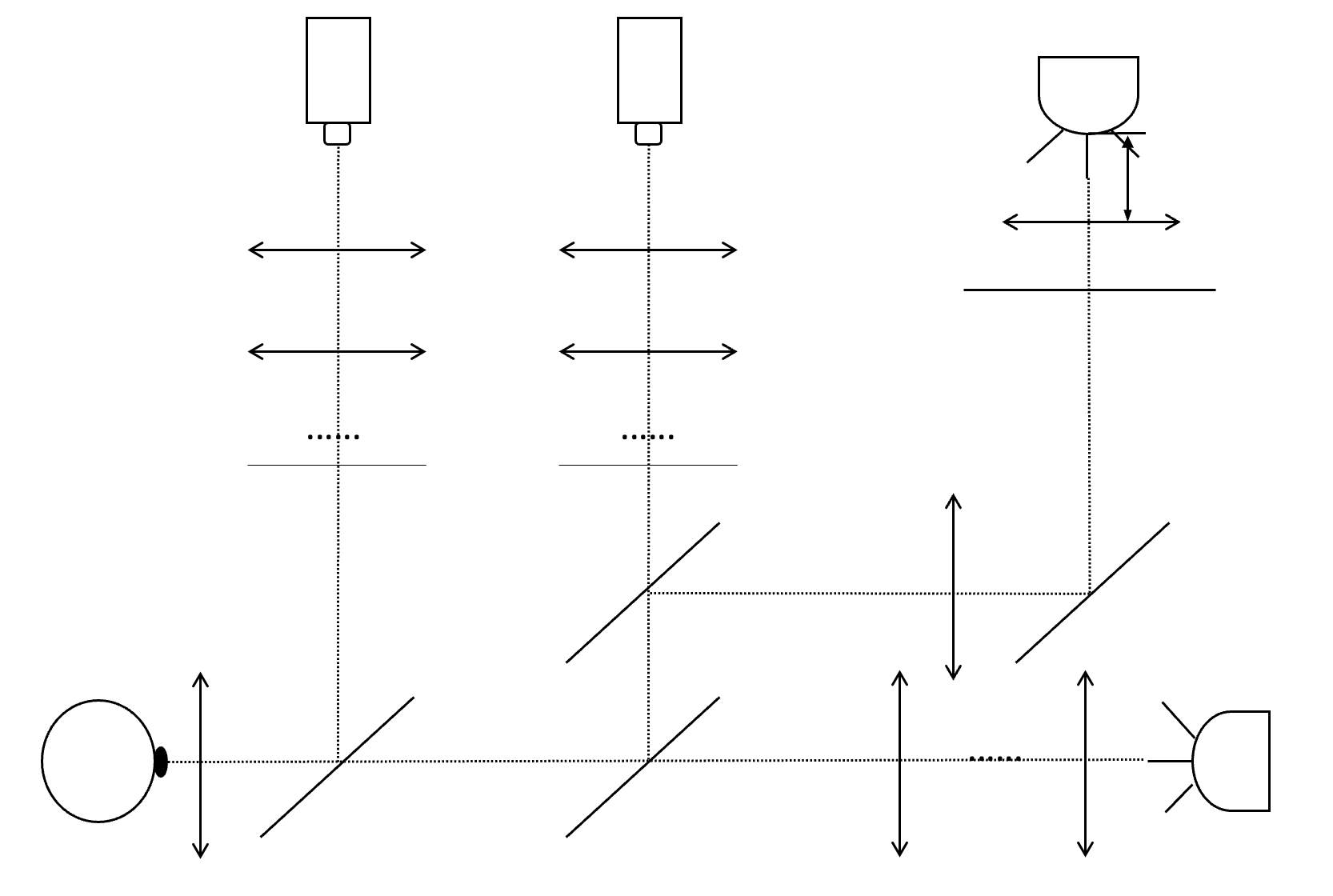}
        \caption{Coupled optical path}
        \label{fig:fig2a}
    \end{subfigure}
    \hfill
    \begin{subfigure}[b]{0.45\textwidth}
        \centering
        \includegraphics[width=\textwidth]{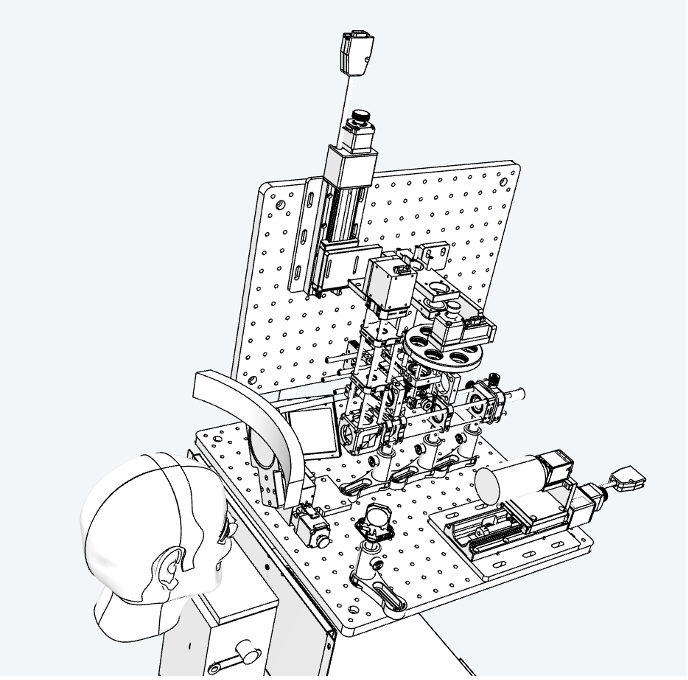}
        \caption{Photomechanical structure}
        \label{fig:fig2b}
    \end{subfigure}
    \caption{Integrated Optical Path for Dual Examination; (a) Coupled optical path; (b) Photomechanical structure}
    \label{fig:fig2}
\end{figure}

\paragraph{U-Net-Based Pupil Localization Method.}
Fundus cameras typically require an additional optical path for pupil detection, which increases system complexity and limits automation. To address this, we employ an infrared illumination–reflection method for pupil localization. The basic principle involves projecting infrared light onto the eye and capturing the reflected image using a camera, with subsequent localization performed algorithmically. In our design, since the fundus imaging system already uses a broadband camera and a 45-degree illumination geometry, pupil localization can be achieved by shifting the chin rest backward so that the pupil is positioned approximately \( d = 25\,\text{mm} \) from the original fundus focal plane. This repurposes the original fundus illumination for the pupil surface. To improve illumination uniformity, the annular aperture is replaced with a full-aperture opening, generating even infrared illumination across the pupil. Based on the lens imaging equation introduced earlier, the required sensor shift is calculated as \( \Delta v = 7.15\,\text{mm} \), ensuring proper focus. By placing a reference object of known dimensions within the field of view, a one-to-one mapping between image coordinates and real-world coordinates can be established for precise pupil localization.

Cross-entropy loss (CEL) is widely used in segmentation tasks due to its simplicity and effectiveness. However, it assumes a balanced class distribution, which is often violated in medical image segmentation such as pupil segmentation. Moreover, CEL is insensitive to boundary localization. The loss is defined as:

\begin{equation}
L = -\frac{1}{N} \sum_{i=1}^N \sum_{c=1}^C y_{i,c} \log \hat{y}_{i,c}
\end{equation}

where \( C \) is the number of classes, \( y_{i,c} \) is the one-hot encoded ground truth, and \( \hat{y}_{i,c} \) is the predicted probability for class \( c \).

To address class imbalance, Dice Loss (DL) is commonly applied. It measures the overlap between prediction and ground truth with class-wise weighting:

\begin{equation}
L_{\text{Dice}} = 1 - \sum_{c=1}^C w_c \cdot \frac{2 \sum_i \hat{y}_{i,c} y_{i,c} + \varepsilon}{\sum_i \hat{y}_{i,c} + \sum_i y_{i,c} + \varepsilon}
\end{equation}

where \( w_c \) is the class weight, typically computed as:

\begin{equation}
w_c = \frac{1}{\sum_i y_{i,c}}
\end{equation}

To improve boundary sensitivity, we incorporate Boundary Aware Loss (BAL), which enhances loss weighting on boundary pixels:

\begin{equation}
L_{\text{boundary}} = \frac{1}{|B|} \sum_{i \in B} \left(1 - \frac{2 \sum \hat{y}_i y_i + \varepsilon}{\sum \hat{y}_i + \sum y_i + \varepsilon} \right)
\end{equation}

where \( B \) denotes the set of boundary pixels, \( \hat{y}_i \) is the predicted probability, and \( y_i \) is the ground truth.

In addition, Surface Loss (SL) minimizes the distance between predicted and ground truth boundaries based on signed distance maps (SDM), allowing sub-pixel optimization:

\begin{equation}
L_{\text{surface}} = \frac{1}{|\Omega|} \sum_{i \in \Omega} \left| S(y_i) - S(\hat{y}_i) \right|
\end{equation}

where \( \Omega \) denotes all pixels in the target region, and \( S(\cdot) \) is the signed distance map function defined as:

\begin{equation}
S(y_i) =
\begin{cases}
-d(y_i, \partial Y), & y_i \in Y \\
+d(y_i, \partial Y), & y_i \notin Y
\end{cases}
\end{equation}

Here, \( d(y_i, \partial Y) \) denotes the Euclidean distance from pixel \( y_i \) to the closest point on the ground truth boundary \( \partial Y \), with the sign indicating inside or outside the target region.

Finally, the total loss integrates all the above components with tunable weights:

\begin{equation}
L = \lambda_1 L_{\text{CEL}} + \lambda_2 L_{\text{BAL}} + \lambda_3 L_{\text{DL}} + \lambda_4 L_{\text{SL}}
\end{equation}

To train a robust pupil segmentation model, we utilized the OpenEDS semantic segmentation dataset~\cite{14}, which consists of 12,759 manually annotated images, including 8,916 for training, 2,403 for validation, and 1,440 for testing. Each image contains four semantic labels: background, sclera, pupil, and iris, making it suitable for our task. To enhance the separability of different eye regions, gamma correction and histogram equalization were applied to the input images. Given the dataset’s sufficient size, no additional data augmentation was applied. For the loss function, the weighting parameters were set as \( \lambda_1 = 1 \), \( \lambda_2 = 10 \), \( \lambda_3 = 1 - \alpha \), and \( \lambda_4 = \alpha \), where \( \alpha \) increases linearly with training epochs and reaches its maximum at half of the total training epochs. The model was trained for 150 epochs using a cosine annealing schedule, where the learning rate decayed from 0.01 to 0.000001. All training was conducted on a server equipped with an NVIDIA GTX 3090 GPU.

\paragraph{Pupil Coordinate Extraction Method Based on Unidirectional Connected Component Analysis.}
After successful pupil segmentation, the next step is to extract the pupil center. Since the pupil and iris may appear non-circular when the eye is not fully aligned with the optical axis, conventional Hough Circle Transform methods often detect multiple candidate circles and centers, leading to ambiguity (see Figure 3a). To address this, we first apply edge detection to extract the contours of ocular surface structures, and then estimate the pupil center based on the geometric center of the segmented region.

To extract edge information, we adopted the classical Canny edge detector, which provides robust performance in identifying object boundaries. The grayscale image is first denoised using a median filter with a kernel size of 3, followed by multi-stage edge detection: gradient calculation, non-maximum suppression, double thresholding, and edge tracking by hysteresis.

The gradient magnitude and orientation are computed using the Sobel operator in both horizontal (\( G_x \)) and vertical (\( G_y \)) directions:

\begin{equation}
M = \sqrt{G_x^2 + G_y^2}, \quad \theta = \arctan \left( \frac{G_y}{G_x} \right)
\end{equation}

To enhance localization accuracy, only local maxima in the gradient direction are preserved, while non-maximum pixels are suppressed. Specifically, a pixel is retained only if its gradient magnitude \( M \) is greater than both neighboring pixels along the direction \( \theta \); otherwise, it is suppressed (set to zero).

Next, two thresholds are applied to classify edge pixels into strong, weak, and non-edge categories. Strong edges are retained directly. Weak edges are preserved only if they are connected to strong edges within an 8-neighborhood; otherwise, they are discarded. This ensures noise suppression and continuity of meaningful contours, as illustrated in Figure 3d. The extracted edges are further used for downstream geometric analysis.

While deep learning combined with edge detection enables effective segmentation of ocular regions, direct edge extraction from raw images is often insufficient. To accurately localize the pupil, we extract the smallest valid connected component by scanning foreground pixels and grouping unvisited neighbors into connected regions. A unidirectional closed contour is identified by tracing edge pixels in a fixed search pattern, rotating directions when necessary, until a closed loop is formed. To reduce false detections, regions with a contour length below 100 pixels are discarded (see Figure 3c).

The pupil center is computed as the centroid of all connected pixels within region \( C \), defined as:

\begin{equation}
cx = \sum_{x_i \in C} x_i, \quad cy = \sum_{y_i \in C} y_i
\end{equation}

As illustrated in Figure 3b, aligning the computed pupil center with the image center enables the system to perform real-time alignment by adjusting the chinrest motor. This fully vision-based approach achieves automated centering with an average processing time of approximately 2.1 seconds, which satisfies practical operation requirements.

\begin{figure}[ht]
    \centering
    \begin{subfigure}[b]{0.45\textwidth}
        \centering
        \includegraphics[width=\textwidth]{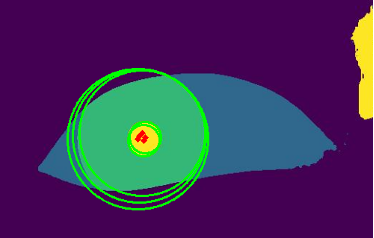}
        \caption{Hough Circle Transform}
        \label{fig:fig3a}
    \end{subfigure}
    \hfill
    \begin{subfigure}[b]{0.45\textwidth}
        \centering
        \includegraphics[width=\textwidth]{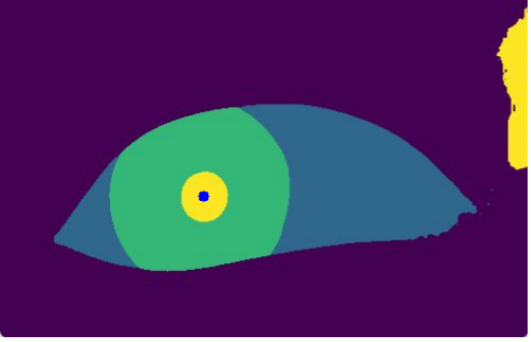}
        \caption{Canny}
        \label{fig:fig3d}
    \end{subfigure}

    \vspace{0.5em}

    \begin{subfigure}[b]{0.45\textwidth}
        \centering
        \includegraphics[width=\textwidth]{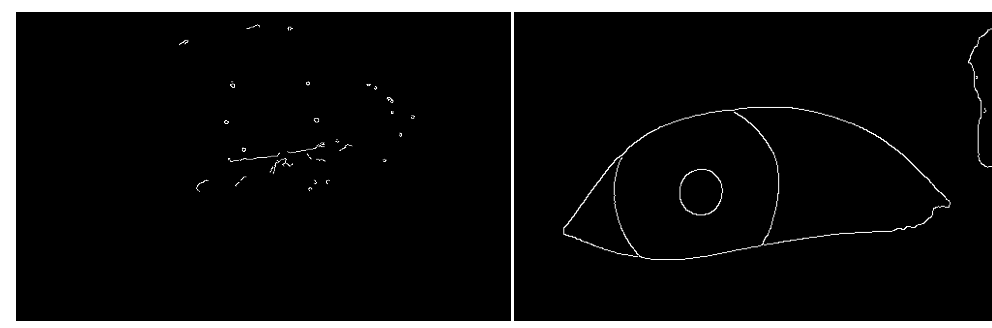}
        \caption{Closed domain extraction}
        \label{fig:fig3c}
    \end{subfigure}
    \hfill
    \begin{subfigure}[b]{0.45\textwidth}
        \centering
        \includegraphics[width=\textwidth]{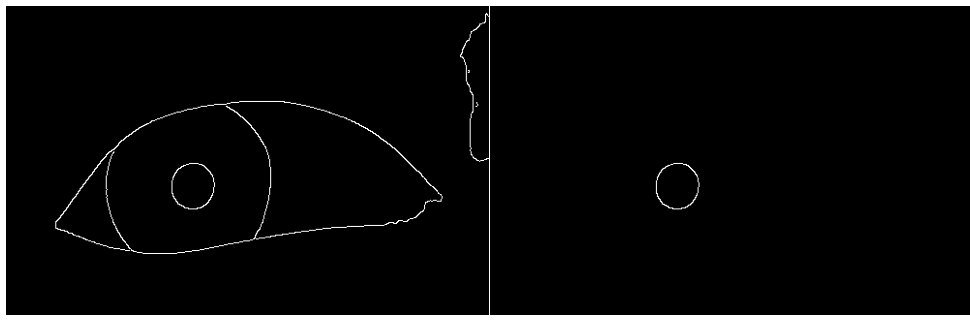}
        \caption{Pupil localization result}
        \label{fig:fig3b}
    \end{subfigure}

    \caption{Extraction of pupil coordinates; (a) Hough Circle Transform; (b) Canny; (c) Unidirectional closed connected domain extraction strategy; (d) Pupil location algorithm results}
    \label{fig:fig3}
\end{figure}

\section{Results}
\label{sec:headings}

\paragraph{Ocular surface structure segmentation.}
Figure 4a shows the pupil image after applying gamma correction and histogram equalization. The first row is the original data, and the second row is the pre-processed image. It can be seen that the pupil and iris are now clearer.
The U-Net model was evaluated using the F1 score and mean Intersection over Union (mIoU) as performance metrics. The model achieved an F1 score of 0.973 and an mIoU of 0.931, demonstrating strong segmentation performance. With a parameter size of only 0.25 million, the model is lightweight and well-suited for deployment on industrial devices to enhance real-time processing. Segmentation results on both raw data and captured model eye images are shown in Figure 4b.

\begin{figure}[ht]
    \centering

    \begin{subfigure}[b]{0.48\textwidth}
        \centering
        \begin{subfigure}[b]{\textwidth}
            \centering
            \includegraphics[width=\textwidth]{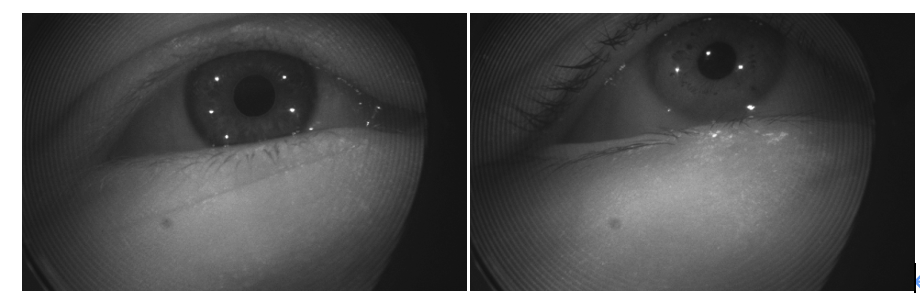}
        \end{subfigure}
        \vspace{0.5em}
        \begin{subfigure}[b]{\textwidth}
            \centering
            \includegraphics[width=\textwidth]{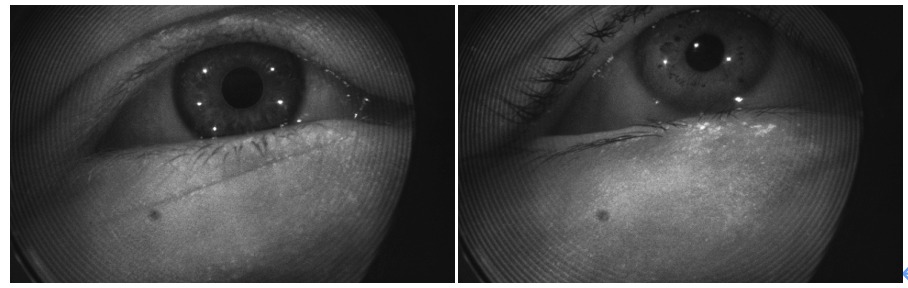}
        \end{subfigure}
        \caption{Preprocessing}
        \label{fig:fig4a}
    \end{subfigure}
    \hfill
    \begin{subfigure}[b]{0.48\textwidth}
        \centering
        \begin{subfigure}[b]{\textwidth}
            \centering
            \includegraphics[width=\textwidth]{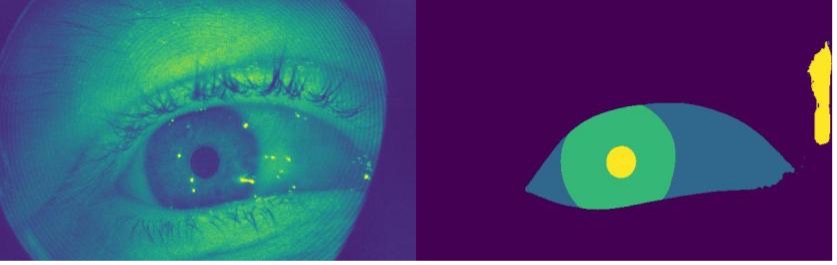}
        \end{subfigure}
        \vspace{0.5em}
        \begin{subfigure}[b]{\textwidth}
            \centering
            \includegraphics[width=\textwidth]{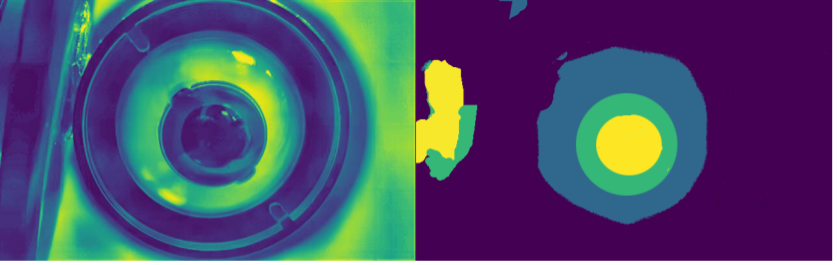}
        \end{subfigure}
        \caption{Segmentation results of validation set and model eyes}
        \label{fig:fig4b}
    \end{subfigure}

    \caption{Pupil image processing; (a) Preprocessing; (b) Segmentation results of validation set and model eyes}
    \label{fig:fig4}
\end{figure}

A comparison was conducted with other lightweight models commonly used for similar tasks, including baseline Res-UNet, DeepLabV3, and MobileUNet. The performance comparison is summarized in Table 1.

\begin{table}[ht]
\centering
\caption{Performance comparison between the improved Dense-UNet and other segmentation models}
\label{tab:model_comparison}
\begin{tabular}{lccc}
\toprule
\textbf{Model} & \textbf{F1} & \textbf{mIoU} & \textbf{Parameters (M)} \\
\midrule
Res-UNet      & 0.952 & 0.913 & 2.56 \\
MobileUNet    & 0.961 & 0.912 & 2.67 \\
DeepLabV3     & 0.969 & 0.927 & 59.33 \\
\midrule
Proposed Model & 0.973 & 0.931 & 0.25 \\
\bottomrule
\end{tabular}
\end{table}

It is worth noting that the network architecture used in this study does not introduce major structural innovations. Both the residual dense connections and the U-Net backbone are based on well-established designs, and the overall performance is expected to be comparable to that of conventional Res-UNet. However, by employing a customized training strategy with a combination of loss functions sensitive to different pixel types, the segmentation performance was effectively improved. This indicates that even without substantial modifications to the network architecture, tailoring the loss function to the characteristics of the specific dataset can enhance model accuracy. Additionally, the use of dense connections allows for parameter reuse, which helps reduce overall model size, making the architecture more suitable for deployment in real-world applications.

\paragraph{Pupil coordinate parameter extraction.}

To evaluate the accuracy of the proposed algorithm, we conducted quantitative analysis on 50 images from the dataset. Expert-annotated pupil centers were used as ground truth, and the Euclidean Distance Error (EDE) between predicted and true centers was computed. We further calculated the Normalized Distance Error (NDE) and the mean squared error (MSE) as follows:

\begin{equation}
EDE = \sqrt{(x_p - x_t)^2 + (y_p - y_t)^2}
\end{equation}

\begin{equation}
NE = \frac{EDE}{r_t}
\end{equation}

\begin{equation}
MSE = \frac{1}{N} \sum_{i=1}^{N} EDE_i^2
\end{equation}

Here, \( N \) is the number of test images, \( (x_p, y_p) \) and \( (x_t, y_t) \) denote the predicted and true coordinates of the pupil center, respectively, and \( r_t \) is the true pupil radius. The results of the quantitative evaluation are presented in Table 2.

\begin{table}[ht]
\centering
\caption{Evaluation results of pupil segmentation accuracy}
\label{tab:pupil_accuracy}
\begin{tabular}{lc}
\toprule
\textbf{Metric} & \textbf{Value} \\
\midrule
EDE (px)              & 2.8 \\
Max EDE (px)          & 4 \\
Min EDE (px)          & 2 \\
Normalized Error (NE) & 0.09 \\
Mean Squared Error (MSE) & 9.1 \\
\bottomrule
\end{tabular}
\end{table}

As shown in the results table, the average localization error is 2.8 pixels, indicating a low positioning deviation. The normalized error is 0.09, meaning the error accounts for approximately 9\% of the pupil radius, demonstrating high detection accuracy. The mean squared error (MSE) is 9.1, reflecting stable error distribution, with the maximum error not exceeding 10\%. Overall, the algorithm achieves high-precision pupil center localization.

In summary, the proposed algorithm enables accurate and rapid extraction of pupil center coordinates. By defining a target coordinate for frontal alignment, the system can automatically adjust the motor to align with the pupil center. Furthermore, the proposed ocular surface segmentation method effectively differentiates between various eye regions. Under frontal-view conditions, additional geometric parameters of the eye surface can be computed, providing supplementary information for downstream analysis.

\paragraph{Fundus photography image quality assessment.}
Before acquiring fundus images with the integrated dual-function device developed in this study, it was necessary to establish a reference baseline image. In the ophthalmology department of Shenzhen University General Hospital, we captured images of the model eye using the Heidelberg fundus camera available there. The resulting color images are shown in the first column of Figure 5a. Following the adjustments described earlier, the clearest fundus image was obtained, as depicted in the second column of Figure 5a. Detailed views of structures such as the nerve fiber layer and vascular branch points are illustrated in Figure 5b.

\begin{figure}[ht]
    \centering

    \begin{subfigure}[b]{0.75\textwidth}
        \centering
        \includegraphics[width=0.48\textwidth]{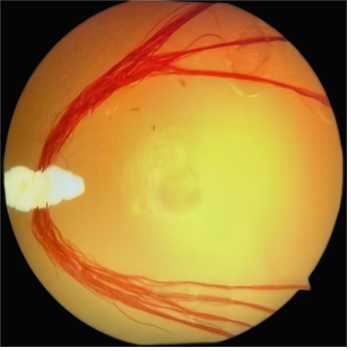}
        \hfill
        \includegraphics[width=0.48\textwidth]{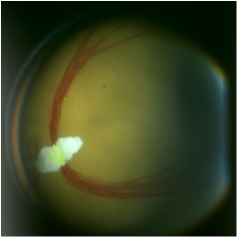}
        \caption{Comparison of fundus blood vessel images in simulated eyes}
        \label{fig:fig5a}
    \end{subfigure}

    \vspace{1.2em}

    \begin{subfigure}[b]{0.5\textwidth}
        \centering
        \includegraphics[width=\linewidth]{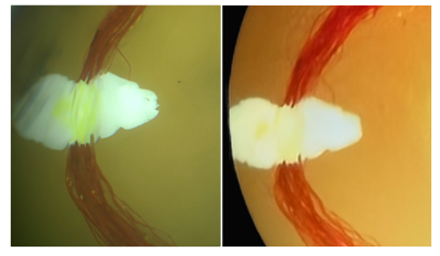}
        \caption{Comparison of image details of ganglia and vascular bifurcations}
        \label{fig:fig5b}
    \end{subfigure}

    \caption{Comparison of imaging using 2-in-1 devices and commercial devices}
    \label{fig:fig5}
\end{figure}

To further assess image quality, we employed several quantitative metrics for evaluating the imaging performance of the fundus illumination system.

First, sharpness was computed using the S3 method described earlier in this work. For contrast evaluation, we adopted the Root Mean Square (RMS) contrast metric, which quantifies the variation in image pixel intensity from the mean brightness. Compared with simpler measures such as maximum-minus-minimum contrast, RMS contrast is more sensitive to overall image detail variation. The RMS contrast is defined as:

\begin{equation}
\bar{I} = \frac{1}{MN} \sum_{x=1}^M \sum_{y=1}^N I_{x,y}
\label{eq:mean_intensity}
\end{equation}

\begin{equation}
C_{\mathrm{RMS}} = \sqrt{ \frac{1}{MN} \sum_{x=1}^M \sum_{y=1}^N (I_{x,y} - \bar{I})^2 }
\label{eq:rms_contrast}
\end{equation}

Here, \( \bar{I} \) represents the mean intensity of the grayscale image. A higher RMS value indicates greater contrast and thus better perceived image sharpness.

The brightness level of the image was also analyzed as a supplementary indicator. In the absence of ambient lighting or interfering display equipment, the mean grayscale value is generally used to estimate global image brightness:

\begin{equation}
\text{Brightness} = \bar{I} = \frac{1}{MN} \sum_{x=1}^M \sum_{y=1}^N I_{x,y}
\label{eq:brightness}
\end{equation}

An overly high brightness value may indicate overexposure or whitening, while an excessively low value may imply underexposure.

For noise level analysis, we adopted the Signal-to-Noise Ratio (SNR) as an evaluation metric. Assuming image brightness fluctuation mainly stems from noise, the SNR is approximated using:

\begin{equation}
\bar{I} = \frac{1}{MN} \sum_{x=1}^M \sum_{y=1}^N I_{x,y}, \quad \sigma = \sqrt{ \frac{1}{MN} \sum_{x=1}^M \sum_{y=1}^N (I_{x,y} - \bar{I})^2 }
\label{eq:snr_components}
\end{equation}

\begin{equation}
\mathrm{SNR(dB)} = 20 \log_{10} \left( \frac{\bar{I}}{\sigma} \right)
\label{eq:snr}
\end{equation}

A higher SNR value indicates a stronger signal and lower noise level.

We compared images captured by our prototype device and a commercial fundus camera. All images were preprocessed by alignment and histogram equalization. Key anatomical structures were extracted using the SIFT method, and rigid transformation was applied for pixel-wise alignment. Global brightness was adjusted consistently prior to quality comparison. Quantitative evaluation results are summarized in Table 3 and Table 4, specifically targeting vessel-rich regions near the optic nerve and ganglia.

\begin{table}[ht]
\centering
\caption{Overall quality evaluation of fundus images}
\label{tab:quality_metrics}
\begin{tabular}{lcc}
\toprule
\textbf{Metric} & \textbf{2-in-1 Device} & \textbf{Commercial Device} \\
\midrule
Spatial Sharpness     & 18.90  & 34.01 \\
Frequency Sharpness   & 3009.36 & 5758.30 \\
Brightness            & 69.87  & 126.37 \\
Contrast              & 35.02  & 65.40 \\
Mean/Variance         & 3.74   & 4.96 \\
\bottomrule
\end{tabular}
\end{table}

\vspace{1em}

\begin{table}[ht]
\centering
\caption{Detail quality evaluation of fundus images}
\label{tab:snr_metrics}
\begin{tabular}{lcc}
\toprule
\textbf{Metric} & \textbf{2-in-1 Device} & \textbf{Commercial Device} \\
\midrule
Spatial Sharpness     & 20.92  & 12.99 \\
Frequency Sharpness   & 6651.25 & 6631.20 \\
Brightness            & 117.27 & 139.70 \\
Contrast              & 42.53  & 54.44 \\
Mean/Variance         & 8.81   & 8.18 \\
\bottomrule
\end{tabular}
\end{table}

Based on the table, the integrated dual-function device successfully achieves basic fundus imaging functionality. However, due to reliance on commercial lenses, high-order aberrations cannot be fully corrected. The compact assembly required additional mirrors compared to the simulation, compromising coaxial alignment and introducing edge-related dispersion. Enhancements such as anti-reflective coatings and increased illumination intensity are needed to improve image quality. Despite these limitations, the device demonstrates superior detail resolution compared to commercial systems, validating the feasibility of the integrated optical design and its potential for further optimization.

\paragraph{Fundus photography image quality assessment.}
Since the current model eye represents an emmetropic (normal) eye, additional measurements were conducted using simulated eyes with known refractive errors, primarily in the myopic range. Commercially available calibration eyes used for refractometer validation were employed, with diopters of –5.25D, –5.25D, and –4.75D. After autofocus was completed using the previously described focusing function, images were captured as shown in Figure 6a. The next step involved extracting and fitting the geometric features of the resulting ring images, starting with skeleton extraction. Traditional skeletonization algorithms derived from the grassfire model perform reasonably well but face challenges when applied to images with varying brightness due to inconsistent reflection intensities caused by different refractive states and high-order aberrations introduced by multiple lenses. These factors limit the effectiveness of fixed and adaptive thresholding in binarization, which in turn hampers accurate skeleton extraction and parameter computation.

To address this, we employed a method derived from the maximal disk transform, combined with an adaptive skeleton point extraction algorithm based on grayscale distribution. After Gaussian denoising and initial binarization, the image centroid was computed using second-order moment features. Rays were then projected from the center to intersect the ring, identifying local maxima in intensity along each ray. Weighted averaging within a defined grayscale range provided coordinates and widths of the intersections. The rays were rotated incrementally, and adaptive thresholding ensured consistent width measurements. Finally, an ellipse fitting process was performed by minimizing the distance constraint from each point to the foci, yielding the ellipse equation and enabling full calculation of the refractive system parameters.
After extracting the geometric parameters from eyes with different refractive powers, the resulting curves are shown in Figure 6b.

The relationship between the refraction value and the ring thickness diameter can be expressed as:

\begin{equation}
R(D) = 0.1136D + 24.4738
\label{eq:refraction_model}
\end{equation}

Refraction detection focuses on the absolute error between the estimated value and the ground truth; therefore, only absolute error is discussed here. The absolute error between the annotated value and the actual value, represented by MAE, is approximately 3--5\%, indicating a degree of discrepancy. As described earlier, the image is still affected by high-order aberrations, leading to suboptimal imaging quality. This suggests the need for further improvements in the optical design. For instance, using aspheric lenses instead of commercial spherical lenses may enhance imaging quality.

From the fitting perspective, the limited number of synthetic eye samples currently involved also constrains the accuracy of the estimation. Nonetheless, the results shown above confirm that the system’s refraction detection functionality works reliably, sufficiently capturing the relationship between changes in brightness and changes in ring thickness diameter, thus validating the theoretical feasibility of the refraction detection method.

\section{Discussion}
Myopia is a common refractive error that, if left uncorrected, may lead to visual fatigue, reduced learning efficiency, and increased risk of high myopia and associated complications such as retinal detachment and macular degeneration. Early detection and scientific intervention are especially critical during adolescence, when refractive status is relatively stable and visual function remains highly adaptable. Beyond adolescents, accurate monitoring of refractive errors and fundus diseases such as diabetic retinopathy, glaucoma, and age-related macular degeneration is equally essential in adults and the elderly to prevent vision loss and irreversible blindness.

Currently, conventional refraction and fundus screening depend heavily on specialized equipment and manual evaluation, resulting in complex workflows and limited scalability, especially in resource-constrained settings. To address these challenges, we proposed an intelligent, integrated device that combines fundus photography and refractive measurement. This dual-function system enables automated, rapid acquisition of both fundus images and refraction data within a short time frame, reducing operational complexity and enhancing accessibility in primary care and underserved areas. The early-warning potential of such a design further supports its public health relevance.

This study presents the following major contributions:
	1.	Based on a comprehensive review of ocular anatomy and optical principles, as well as existing fundus and refraction systems, we developed a prototype that integrates both modalities using wavelength-based optical separation and electromechanical design. The system successfully merges both imaging paths into a single optical axis, and its performance was comprehensively evaluated.
	2.	To enhance automation, we introduced a Dense-U-Net-based segmentation network that achieves accurate segmentation of ocular surface structures. Combined with edge detection and unidirectional connected component extraction, the method enables rapid localization of the pupil boundary and center. This allows for automatic alignment of the optical axis using motorized control. Furthermore, a passive autofocus strategy was implemented, and the sharpness at multiple focal positions was assessed using the S3 index. A focus-position function was fitted for both refraction and fundus imaging paths, improving focusing efficiency.

Despite these achievements, several limitations remain. Although lens parameters were optimized during simulation, the prototype still relies on commercial lenses, resulting in some degradation in image quality. High-order aberrations were not fully analyzed or corrected, and stray light suppression remains incomplete. Due to hardware constraints, high-resolution training for the segmentation network was not implemented, and no animal or clinical testing has been conducted to date. These issues will require further optimization. Nonetheless, the proposed device and its supporting algorithms have achieved the primary design goals and demonstrate strong potential as a dual-function ophthalmic screening system.

Future improvements will focus on three main aspects. First, optical and mechanical hardware optimization is needed. Custom lens design, including correction of high-order aberrations and implementation of aspheric lenses, will enhance imaging resolution and signal-to-noise ratio. A more compact and portable design with lightweight materials will improve usability in community health centers, school screenings, and rural areas. Second, clinical validation should be expanded through large-scale studies in real-world healthcare environments. Collaboration with medical institutions and public health agencies could promote application in myopia prevention and early detection of diabetic retinopathy. Integration with IoT and cloud computing will enable remote storage, intelligent analysis, and personalized eye health management.

Third, intelligent interaction and human-factor optimization should be pursued. Incorporating voice-guided instructions and AI-driven automation can lower the operational barrier for non-specialist users. In addition, augmented reality (AR) and visual interfaces may assist physicians in diagnostic decision-making, positioning the device as not only a screening tool but also a valuable clinical aid in complex cases.

\section{Conclusion}

This study presents a novel, integrated device that combines fundus photography and refractive measurement into a compact, automated system for ophthalmic screening. The proposed solution addresses key limitations in current practices by enabling fast, accurate, and user-friendly examinations, with significant potential for deployment in diverse clinical and public health contexts.

With further optimization of optical design, enhancement of deep learning algorithms, expanded clinical validation, and integration of intelligent features, this dual-function system is poised to become a practical and scalable tool for early diagnosis and long-term monitoring of ocular health. It offers a promising avenue for improving vision care accessibility and supporting the development of personalized, technology-driven eye health management.

\end{document}